\documentclass[aps,eqsecnum,nofootinbib,superscriptaddress,showpacs]{revtex4}
\usepackage[dvips]{graphicx}  
\usepackage{amsmath,amsfonts}
\usepackage{bm}
\usepackage{color}  
\def\be{\begin{eqnarray}}
\def\ee{\end{eqnarray}}
\def\beq{\begin{equation}}
\def\eeq{\end{equation}}

\def\p{\partial}

\def\({\left (}
\def\){\right )}
\def\[{\left [}
\def\[{\right ]}

\newcommand{\bra}[1]{\big\langle\, #1\, \big\vert}
\newcommand{\ket}[1]{\big\vert\, #1\, \big\rangle}
\newcommand{\IP}[2]{\big\langle\, #1\, \big\vert\, #2\, \big\rangle}
\newcommand{\Exp}[1]{\langle\, #1\, \rangle}
\newcommand{\GL}{Ginzburg-Landau }
\newcommand{\AdS}{\text{AdS}}
\bmdefine{\bmk}{\bm{k}}
\bmdefine{\bmx}{\bm{x}}
\bmdefine{\bmA}{\bm{A}}
\bmdefine{\bmB}{\bm{B}}
\bmdefine{\bmJ}{\bm{J}}
\newcommand{\hpsi}{\Hat{\psi}}
\newcommand{\calL}{\mathcal{L}}
\newcommand{\calO}{\mathcal{O}}
\newcommand{\tilk}{\Tilde{k}}

\newcommand{\tilPhi}{\Tilde{\Phi}}
\newcommand{\tilPsi}{\Tilde{\Psi}}
\newcommand{\odiff}[2]{ \frac{d #1}{d #2} }
\newcommand{\odiffII}[2]{ \frac{d^2 #1}{d #2^2} }
\begin{document}
\title{Characteristic length of an AdS/CFT superconductor}
\author{Kengo Maeda}
\email{maeda302@sic.shibaura-it.ac.jp}
\affiliation{Department of Engineering,
Shibaura Institute of Technology, Saitama, 330-8570, Japan}

\author{Takashi Okamura}
\email{tokamura@kwansei.ac.jp}
\affiliation{Department of Physics, Kwansei Gakuin University,
Sanda, 669-1337, Japan}

\date{\today}
\begin{abstract}
We investigate in more detail the holographic model
of a superconductor recently found by Hartnoll, Herzog, and
Horowitz~[Phys.\ Rev.\ Lett.\ {\bf 101}, 031601],
which is constructed from a condensate of a charged scalar field
in $\AdS_4$-Schwarzschild background.
By analytically studying the perturbation of the gravitational system
near the critical temperature $T_c$, we obtain
the superconducting coherence length proportional
to $1/\sqrt{1-T/T_c}$ via AdS/CFT correspondence.
By adding a small external homogeneous magnetic field to the system,
we find that a stationary diamagnetic current proportional
to the square of the order parameter is induced by the magnetic field.
These results agree with \GL theory
and strongly support the idea that a superconductor
can be described by a charged scalar field on a black hole
via AdS/CFT duality.
\end{abstract}
\maketitle
\section{Introduction}\label{sec:intro}
The AdS/CFT duality \cite{maldacena97} gets into the limelight
as a powerful tool to investigate the strongly coupled gauge theories.
Motivated by the recent experimental data of quark-gluon plasma
at the Relativistic Heavy Ion Collider \cite{Shuryak04,Shuryak05},
transport coefficients such as shear viscosity were
calculated for various duality models.
Interestingly, for a large class of dualities,
it was found that the ratio of viscosity divided by entropy density
is a universal constant compatible with the experimental data~
(see, for example, Refs. \cite{KovSonStar04,mno06} for all references).
This leads us to expect that strongly coupled phenomena
such as quantum phase transition or superconducting phase transition
in condensed matter systems are described by some kind of duality.

If a superconductor can be described by a gravitational model
via the AdS/CFT duality, there should be a scalar hair on
an anti-deSitter (AdS) black hole
which represents the condensation in the dual gauge theory.
Gubser \cite{gubser89}
has presented a counter example to a no scalar hair theorem by giving
a static solution of a charged scalar field coupled to
an Abelian gauge field on the $\AdS_4$-Reissner-Nortstr\"om
black hole background if the charge of the black hole is large enough.
Since the temperature of the black hole decreases
as the charge increases, the black hole can support
the scalar hair only for low temperature.
This is because the scalar field condensation
breaks the Abelian gauge symmetry spontaneously
at sufficiently low temperature.
Hartnoll et. al. \cite{HHH2008} numerically showed that
there is a critical temperature below
which the charged scalar hair exists on
$\AdS_4$-Schwarzschild black hole and the conductivity becomes
infinite at the low frequency limit.
It was also numerically shown that
the scalar field condensation occurs
below a critical temperature under the presence
of an external magnetic field \cite{nakanowen,albashjohnson,wen}.
Quite recently, general properties of p-wave superconductors
were investigated by Gubser and Pufu \cite{gubserpufu60}
and independently by Roberts and Hartnoll \cite{robertshartnoll98}
in a model of a non-Abelian gauge field
in the background of $\AdS_4$-Schwarzschild black hole.

The purpose in this paper is to explore a little further
the model of the superconductor composed of the charged scalar field
on $\AdS_4$-Schwarzschild background \cite{HHH2008}
by investigating perturbation of the system
near the critical temperature.
The order parameter of the superconductor is the scalar operator
dual to the charged scalar field.
So, the correlation length of the order parameter, or
the superconducting coherence length $\xi$ is obtained
by the perturbation of the scalar field.
According to \GL theory,
a superconductor is characterized by only two parameters,
$\xi$ and the magnetic penetration length $\lambda$.
Therefore, it will be of interest to determine
the two parameters by investigating the perturbation.
Motivated by this, we analytically investigate static fluctuation
of the scalar field with nonzero spatial momentum
along one spatial coordinate of the AdS boundary
to obtain the superconducting coherence length $\xi$
via AdS/CFT correspondence.
Following \cite{HHH2008}, we take the probe limit
where the fluctuation do not backreact
on the original $\AdS_4$-Schwarzschild geometry.
Under the probe limit we also investigate static fluctuation
of the Abelian gauge field forming a homogeneous magnetic field
as a first step to derive the magnetic penetration length $\lambda$.

The plan of our paper is as follows: In section II
the charged scalar field solution obtained in the holographic model
\cite{HHH2008} is reconstructed by perturbation technique.
In section III we derive $\xi$ by analyzing
the equations for the perturbation.
In section IV we observe that the diamagnetic current
can be induced by the small homogeneous magnetic field.
Section V is devoted in conclusions and discussion.
\section{A model of superconductor in AdS/CFT}
In this section we reconstruct the charged scalar field solution
numerically obtained in \cite{HHH2008}
by using the regular perturbation theory technique.
The background spacetime is $\AdS_4$-Schwarzschild black hole
with the metric
\begin{align}
  & ds^2
  = - f(r)\, dt^2 + \frac{dr^2}{f(r)} + \frac{r^2}{L^2}\, (dx^2+dy^2)~,
& & f(r) = \frac{r^2}{L^2}\, \left( 1 - \frac{r_0^3}{r^3} \right)~,
\label{Sch-eq}
\end{align}
where $r_0$ is the horizon radius and $L$ is the AdS radius.
The Hawking temperature $T$ is given by
\be
  T = \frac{3}{4 \pi}\, \frac{r_0}{L^2}~.
\ee

It is convenient to introducing a new coordinate $u := r_0/r$,
and the metric (\ref{Sch-eq}) is written as
\begin{align}
  & ds^2
  = \frac{L^2\, \alpha^2(T)}{u^2}\,
    \left( - h(u)\, dt^2 + dx^2 + dy^2 \right)
  + \frac{L^2\, du^2}{u^2\, h(u)}~,
& & h(u) = 1 - u^3~,
\label{eq:metric}
\end{align}
where $\alpha(T) := 4 \pi\, T/3 = r_0/L^2$.

We consider the matter fields on the background spacetime which
consist of a Maxwell field and a charged complex scalar field
with charge $e$ and mass $m$.
The Lagrangian density is given by
\begin{align}
  & {\cal L}
  = \frac{L^2}{2 \kappa_4^2\, e^2}\, \sqrt{-g}\, \left(
  - \frac{1}{4}\, F^{\mu\nu} F_{\mu\nu}
  - |D \Psi\,|^2 - m^2\, |\Psi|^2 \right)~,
& & D_\mu := \partial_\mu - i\, A_\mu~.
\label{action}
\end{align}

Following \cite{HHH2008}, we shall confine our attention
to the case $L^2\, m^2 = -2$
and consider the probe limit in which the gauge field and scalar field
do not backreact on the original metric (\ref{Sch-eq}).
This limit is realized by taking the limit $e\to \infty$,
keeping $A_\mu$ and $\Psi$ fixed.
So, the equations of motion for $A_\mu$ and $\Psi$ are decoupled
from Einstein's equations and we obtain the following equations
\begin{align}
  & 0 = D^2\, \Psi - m^2\, \Psi~,
& & 0
  = \nabla_\nu F_\mu{\,}^{\nu}
  - i\, \left[ \left( D_\mu \Psi \right)^\dagger\, \Psi
    - \Psi^\dagger \left( D_\mu \Psi \right) \right]~.
\label{eq:basic-eq}
\end{align}

Under the ansatz
\begin{align}
  & \Psi=\Psi(u)~,
& & A_\mu = \Phi(u)\, (dt)_\mu~,
\label{eq:bg_ansatz}
\end{align}
the equations of motion (\ref{eq:basic-eq}) are reduced to
\begin{align}
  & 0 = \left( u^2 \odiff{}{u}\, \frac{h(u)}{u^2}\, \odiff{}{u}
  - \frac{L^2\, m^2}{u^2} \right)\, \tilPsi
  + \frac{\tilPhi^2}{h(u)}\, \tilPsi~,
\label{eq:bg-scalar_eq} \\
  & 0 = h(u)\, \odiffII{\tilPhi}{u}
  - \frac{2\, \vert\, \tilPsi\, \vert^2}{u^2}\, \tilPhi~.
\label{eq:bg-EM_eq}
\end{align}
where
new variables $\tilPhi := \Phi/\alpha(T)$ and $\tilPsi := L\, \Psi$
are dimensionless quantities.
Without loss of generality, we can set $\tilPsi$ to be real.

The trivial solution is easily found as
\begin{align}
  & \tilPsi = 0~,
& &
  \tilPhi
  = \mu/\alpha(T) - q\, u
  = q\, ( 1 - u )~,
\label{trivial}
\end{align}
where $\mu$ is interpreted as the external source
in the dual $(2+1)$-dimensional gauge theory
and it is determined by the condition $A_\mu dx^\mu$
to be well-defined at the horizon, i.e., $\Phi(u=1) = 0$
\cite{ref:KobayashiEtAL}.
Dimensionless constant $q$ is related to
the dual charge density coupled to $\mu$ as,
\begin{align}
  & \Exp{J^t(x)}
  = \left.
    \frac{ \delta S_{\text{on-shell boundary}} }{\delta A_t(x)}\,
    \right\vert_{u=0}
  = \frac{L^2}{2 \kappa_4^2\, e^2}\,
    \left( \frac{4 \pi \, T}{3} \right)^2\, q~.
\label{eq:R_charge_density}
\end{align}

The non-trivial solution asymptotically behaves
near the AdS boundary as
\begin{align}
  & \tilPsi
  = \tilPsi^{(-)}\, u^{\Delta_-} + \tilPsi^{(+)}\, u^{\Delta_+}
  + \cdots~,
& & \tilPhi = \mu/\alpha(T) - q\, u + \cdots~,  
\end{align}
where $\Delta_{\pm} := (3 \pm \sqrt{9 + 4\, L^2\, m^2})/2$.
For $L^2\,m^2=-2$ case, we obtain $\Delta_- = 1$ and $\Delta_+ = 2$, 
where both falloffs of $\tilPsi$ are normalizable.  

Each coefficient $\tilPsi^{(\pm)}$ is proportional
to the condensate thermal expectation value of the scalar operator
$\Exp{\calO_\pm}$ of dimension $\Delta_{\pm}$.
To obtain a stable solution,
we must impose either $\tilPsi^{(-)}=0$ or $\tilPsi^{(+)}=0$.
So, the asymptotic boundary condition of the scalar field
$\tilPsi$ dual to the scalar operator $\Exp{\calO_-}$ ($\Exp{\calO_+}$)
is $\tilPsi^{(+)}=0$ ($\tilPsi^{(-)}=0$), and
$\tilPsi$ has the asymptotic behavior near the AdS boundary as
\begin{align}
  & \tilPsi = \tilPsi^{(I)}\, u^{\Delta_I}\, \big[~1 + O(u^2)~\big]~,
\label{eq:bc-bg_Psi}
\end{align}
where $I = \pm$ for $\Exp{\calO_\pm}$.

Since the trivial solution $\tilde{\Phi}$
in Eqs.(\ref{eq:bg-scalar_eq}) and (\ref{eq:bg-EM_eq})
is parametrized by the dimensionless constant
$q\propto \Exp{J^t}/T^2$ only,
the non-trivial solution $\tilde{\Psi}$ emerges
above a critical value $q_c$ under the boundary condition.
According to the numerical calculation~\cite{HHH2008}, the thermal
expectation value $\Exp{\calO_I}$ behaves as
\begin{align}
  & \tilPsi^{(I)} \propto \Exp{\calO_I}
  \sim \left( 1 - T/T_c \right)^{1/2}
\label{o-behavior}
\end{align}
for a given $\mu$~(or $\Exp{J^t}$), or equivalently
\be
\Exp{\calO_I} \sim \left( q/q_c - 1 \right)^{1/2}
\label{o-behavior1}
\ee
near the critical temperature $T_c$.
In the limit $T\to T_c$, $\epsilon:=q/q_c-1~(>0)$ is a small parameter,
and the non-trivial solution to Eqs.(\ref{eq:bg-scalar_eq})
and (\ref{eq:bg-EM_eq}) can be obtained
as a series in $\epsilon$.
From the continuity, the solution at the critical temperature
should be
\begin{align}
  & \tilPsi_c = 0~,
& & \tilPhi_c = q_c\, ( 1 - u )~.
\end{align}
So, we can expand $\tilde{\Psi}$ and $\tilde{\Phi}$ as
\begin{align}
  & \tilPsi(u)
  = \epsilon^{1/2}\, \tilPsi_1(u)
  + \epsilon^{3/2}\, \tilPsi_2(u) + \cdots~,
& & \tilPhi(u)
  = \tilPhi_c(u) + \epsilon\, \tilPhi_1(u) + \cdots~.
\label{back-solution}
\end{align}
Here, we should note that the difference of $\epsilon$-behavior
between $\tilPsi$ and $\tilPhi$ comes from Eqs.(\ref{eq:bg-scalar_eq})
and (\ref{eq:bg-EM_eq}).

Substituting Eq.(\ref{back-solution}) into
Eqs.(\ref{eq:bg-scalar_eq}) and (\ref{eq:bg-EM_eq})
we obtain equations for $\tilPsi_1$ and $\tilPhi_1$:
\begin{align}
  & 0 = {\cal L}_\psi\, \tilPsi_1~,
& & 0
  = \odiffII{\tilPhi_1}{u}
  - \frac{2\, \vert\, \tilPsi_1\, \vert^2\, \tilPhi_c}
         {u^2\, h(u)}~,
\label{eq:pert-bg-EM_eq}
\end{align}
where the differential operators ${\cal L}_\psi$ is defined by
\begin{align}
  & {\cal L}_\psi
  := - \left( u^2\, \odiff{}{u}\, \frac{h(u)}{u^2}\, \odiff{}{u}
  - \frac{L^2\, m^2}{u^2}
  + \frac{\tilPhi_c^2}{h(u)} \right)~.
\label{diffe-op}
\end{align}
By imposing the regularity condition at the horizon
\begin{align}
  & \left. \frac{1}{\tilPsi_1}\, \odiff{\tilPsi_1}{u}\,
    \right\vert_{u=1}
  = - \frac{L^2\, m^2}{3} = \frac{2}{3}~,
\end{align}
we find the constant $q_c$ for which there is a unique regular solution
satisfying the asymptotic boundary condition mentioned above.
In $I=+$ case, for example, $q_c \sim 4.07$,
which is consistent with the numerical result in \cite{HHH2008}.
\section{Superconducting coherence length}
In this section,
we will determine the superconducting coherence length $\xi$
by investigating fluctuations
around the background field (\ref{eq:bg_ansatz})
$(\tilPsi(u), \tilPhi(u))$.
It is enough to consider static perturbations for the purpose,
so let us confine our attention to the fluctuations
with only spatial momentum along $x$-direction:
\begin{align}
  & \delta A_\mu(u, x)\, dx^\mu
  = \Big[~A_x(u, k)\, dx + A_y(u, k)\, dy + \phi(u, k)\, dt~
    \Big]\, e^{i k x}~,
\nonumber \\
  & \delta\Psi(u, x)
  = \frac{1}{L\, \alpha(T)}\,
  \Big[~\psi(u, k) + i\, \hpsi(u, k)~\Big]\, e^{i k x}~,
\label{perturbation}
\end{align}
where both functions $\psi$ and $\hpsi$ are real and metric 
fluctuations of the order of the gauge and scalar fluctuations can be 
consistently set to zero under the probe limit.
From the perturbed equations derived from Eq.(\ref{eq:basic-eq}),
we find the following three linearized equations
for $\phi$, $\psi$, and $A_y$ decoupled from the other variables:
\begin{align}
  & \tilk^2\, \psi
  = \left( u^2\, \odiff{}{u}\, \frac{h(u)}{u^2}\, \odiff{}{u}
  - \frac{L^2\, m^2}{u^2}
  + \frac{\tilPhi^2}{h(u)} \right)\, \psi
  + \frac{2\, \tilPhi\, \tilPsi}{h(u)}~\phi~,
\label{psi-eq} \\
  & \tilk^2\, \phi
  = \left( h(u) \odiffII{}{u}
  - \frac{2\, \tilPsi^2}{u^2} \right)\, \phi
  - \frac{4\, \tilPhi\, \tilPsi}{u^2}\, \psi~,
\label{st-eq} \\
  & \tilk^2\, A_y
  = \left( \odiff{}{u}\, h(u)\, \odiff{}{u}
  - \frac{2\, \tilPsi^2}{u^2} \right)\, A_y~,
\label{sy-eq}
\end{align}
where $\tilk := k/\alpha(T)$.

The superconducting coherence length $\xi$ is nothing but
the correlation length of the order parameter,
and $\xi$ appears as the pole of the static correlation function
of the order parameter in the Fourier space
\begin{align}
  & \Exp{\Tilde{\calO}(\Vec{k}\,) \Tilde{\calO}(-\Vec{k}\,)}
  \sim \frac{1}{\vert\, \Vec{k}\, \vert^2 + 1/\xi^2}~.
\end{align}
Since the complex scalar field $\Psi$ plays a role
of the order parameter in our model
and the background $\tilPsi$ is real,
the real part of the order parameter fluctuation $\psi$ gives
the superconducting coherence length.

The pole of the static correlation function of a dual field operator
is obtained
by solving the eigenvalue problem for the static perturbation
with wave number $k$ of the corresponding bulk field
as $1/\xi^2 = - k_*^2$,
where $k_*$ is a wave number permitted as eigenvalues.
In the present case, our task is to evaluate
eigenvalues $\tilk^2$ for Eqs.(\ref{psi-eq}) and (\ref{st-eq})
under the appropriate boundary conditions.

Since it is difficult to solve the eigenvalue equations
(\ref{psi-eq}) and (\ref{st-eq}) analytically,
we solve the equations as a series in $\epsilon$
near the critical temperature $T_c$.
According to \GL theory, $\xi$ diverges to infinity as $T \to T_c$.
This implies that there exists zero eigenvalue $k_* = 0$ solution
at the critical temperature $T_c$.
Hereafter, we shall confine our attention to the eigensystem
with $\lim_{\epsilon \to 0}~k_* = 0$.

From the behavior (\ref{back-solution}),
Eqs.(\ref{psi-eq}) and (\ref{st-eq}) are expanded in $\epsilon$ as
\begin{align}
  & - \tilk^2\, \psi
  = \left( \calL_\psi
    - \frac{2\, \epsilon \tilPhi_c\, \tilPhi_1}{h} \right)\, \psi
  - \frac{2\, \epsilon^{1/2}\, \tilPhi_c\, \tilPsi_1}{h}~\phi~,
\label{psi-tildepsi} \\
  & - \tilk^2\, \phi
  = \left( - h\, \odiffII{}{u}
    + \frac{2\, \epsilon\, \tilPsi_1^2}{u^2} \right)\, \phi
  + \frac{4\, \epsilon^{1/2}\, \tilPhi_c\, \tilPsi_1}{u^2}\, \psi~.
\label{phi-tildepsi}
\end{align}

The boundary conditions for Eqs.(\ref{psi-tildepsi})
and (\ref{phi-tildepsi}) are as follows:
at the horizon,
\begin{align}
  & \psi(1) = \text{regular}~,
& & \phi(1) = 0~,
\label{eq:horizon_bc}
\end{align}
and near the AdS boundary
\begin{align}
  & \psi(u)
  = (\text{const.}) \times u^{\Delta_I} \big[~1 + O(u^2)~\big]~,
& & \phi(u)
  = (\text{const.}) \times u + O(u^2)~.
\label{eq:infinity_bc-phi}
\end{align}

In the eigenvalue equations (\ref{psi-tildepsi})
and (\ref{phi-tildepsi}) with the boundary conditions
(\ref{eq:horizon_bc}) and (\ref{eq:infinity_bc-phi}),
the infinitesimal expansion parameter is $\epsilon^{1/2}$, so
one may expect that we have eigenvalue $\tilk_*^2 = O(\epsilon^{1/2})$.
However, we have $\tilk_*^2 = O(\epsilon)$ as seen later.

It is easy to show that the zeroth order solution,
$\psi_0$ and $\phi_0$, for the eigenvalue equation
(\ref{psi-tildepsi}) and (\ref{phi-tildepsi})
satisfying the boundary conditions (\ref{eq:horizon_bc}) and
(\ref{eq:infinity_bc-phi}) is given by
\begin{align}
  & \psi_0 = \tilPsi_1~,
& & \phi_0 = 0~,
\end{align}
where we use $\calL_\psi\, \tilPsi_1 = 0$.
This means that $\phi = O(\epsilon^{1/2})$ at most
from Eq.(\ref{phi-tildepsi}).
So we put $\phi =: \epsilon^{1/2}\, \varphi$, and
Eqs.(\ref{psi-tildepsi}) and (\ref{phi-tildepsi}) are rewritten by
\begin{align}
  & - \tilk^2\, \psi
  = \calL_\psi\, \psi
  - \epsilon\, \frac{2\, \tilPhi_c}{h}
    \left( \tilPhi_1\, \psi + \tilPsi_1\, \varphi \right)~,
\label{eq:pert_eq-psi} \\
  & - \tilk^2\, \varphi
  = \left( - h\, \odiffII{}{u}\, \varphi
    + \frac{4\, \tilPhi_c\, \tilPsi_1}{u^2}\, \psi \right)
  + \epsilon\, \frac{2\, \tilPsi_1^2}{u^2}\, \varphi~.
\label{eq:pert_eq-phi}
\end{align}
Thus, the expansion parameter of R.H.S.
in Eqs.(\ref{eq:pert_eq-psi}) and (\ref{eq:pert_eq-phi}) is $\epsilon$,
implying $\tilk_*^2 = O(\epsilon)$.
The temperature dependence of the coherence length,
$\xi \propto (-\tilk_*^2)^{-1/2}
\propto ( 1 - T/T_c )^{-1/2}$,
is an expected behavior in \GL theory.

Now let us evaluate the superconducting coherence length $\xi$
at the leading order in $\epsilon$.
We expand $\psi$, $\varphi$, and $\tilk_*^2$ as
\begin{align}
  & \psi = \tilPsi_1 + \epsilon\, \psi_1 + O(\epsilon^2)~,
& & \varphi = \varphi_0 + O(\epsilon)~,
& & \tilk_*^2 = \epsilon\, \big( \tilk^2 \big)_1 + O(\epsilon^2)~.
\end{align}
Then, Eq.(\ref{eq:pert_eq-psi}) is rewritten up to $O(\epsilon)$ as
\begin{align}
  & - \big( \tilk^2 \big)_1\, \tilPsi_1
  = \calL_\psi\, \psi_1
  - \frac{2\, \tilPhi_c\, \tilPsi_1}{h}
    \left( \tilPhi_1 + \varphi_0 \right)~,
\label{eq:pert_eq-psiII}
\end{align}
and the equation of motion for $\varphi_0$ is given by
\begin{align}
  & \odiffII{\varphi_0}{u}
  = \frac{4\, \tilPhi_c\, \tilPsi_{1}^2}{u^2\, h}
  = 2\, \odiffII{\tilPhi_1}{u}~,
\label{eq:zeroth_eq-etaII}
\end{align}
where we use Eq.(\ref{eq:pert-bg-EM_eq}).
The solution of Eq.(\ref{eq:zeroth_eq-etaII})
with the boundary conditions (\ref{eq:horizon_bc})
and (\ref{eq:infinity_bc-phi}) is given by
\begin{align}
  & \varphi_0(u)
  = 2 \left[~\tilPhi_1(u) - \tilPhi_1(0)\, (1 - u)~\right]
  \in \mathbb{R}~,
\label{eq:sol_phi_1}
\end{align}

For states $\psi_I$, $\psi_{II}$ with the boundary conditions
(\ref{eq:horizon_bc}) and (\ref{eq:infinity_bc-phi}),
let us introduce an inner product
\begin{align}
   \IP{\psi_I}{\psi_{II}}
  &:= \int_0^1 \frac{du}{u^2}~\psi_I^*(u)~\psi_{II}(u)~.
\label{eq:def-IP}
\end{align}
It is easily checked that $\calL_\psi$ is hermitian
for the inner product (\ref{eq:def-IP}).

Making use of $\calL_\psi\, \tilPsi_1 = 0$ and
hermiticity of $\calL_\psi$,
the inner product between $\tilPsi_1$ and
Eq.(\ref{eq:pert_eq-psiII}) gives us
\begin{align}
   - \big( \tilk^2 \big)_1\, \IP{\tilPsi_1}{\tilPsi_1}
  &= \bra{\tilPsi_1}\, \calL_\psi\, \ket{\psi_1}
  - \bigg\langle\, \tilPsi_1\,
    \bigg\vert\,
       \frac{2\, \tilPhi_c\, \tilPsi_1}{h}
       \left( \tilPhi_1 + \varphi_0 \right)\,
    \bigg\rangle
\nonumber \\
  &=- \bigg\langle\, \tilPsi_1\,
    \bigg\vert\, \frac{2\, \tilPhi_c\, \tilPsi_1}{h}\, \tilPhi_1\,
    \bigg\rangle
  - \int^1_0 du~
    \frac{2\, \tilPhi_c\, \tilPsi_1^2}
         {u^2\, h}\, \varphi_0
\nonumber \\
  &=- \bigg\langle\, \tilPsi_1\,
    \bigg\vert\, \frac{2\, \tilPhi_c\, \tilPsi_1}{h}\, \tilPhi_1\,
    \bigg\rangle
  + \frac{1}{2} \int^1_0 du~\left( \odiff{\varphi_0}{u} \right)^2~,
\label{eq:IP-tilPsi_1-pert_eq}
\end{align}
where we used Eq.(\ref{eq:zeroth_eq-etaII}) and the boundary conditions
(\ref{eq:horizon_bc}) and (\ref{eq:infinity_bc-phi})
in the third equality.

We can show that the first term in Eq.(\ref{eq:IP-tilPsi_1-pert_eq})
vanishes as follows:
From Eq.(\ref{eq:bg-scalar_eq}), we have the equation of motion
for $\tilPsi_2$ defined by Eq.(\ref{back-solution}) as
\begin{align}
  & \calL_\psi\, \tilPsi_2
  = \frac{2\, \tilPhi_c\, \tilPsi_1}{h}\, \tilPhi_1~,
& & \tilPsi_2(1) = \text{regular}~,
& & \tilPsi_2(0) = (\text{const.}) \times u^{\Delta_I}~.
\label{eq:EOM-tilPsi_2}
\end{align}
So the inner product (\ref{eq:def-IP}) is well-defined
for $\tilPsi_2$, and Eq.(\ref{eq:EOM-tilPsi_2}) gives us
\begin{align}
   0
  &= - 2\, \IP{\calL_\psi\, \tilPsi_1}{\tilPsi_2}
  = - 2\, \IP{\tilPsi_1}{\calL_\psi\, \tilPsi_2}
  = - 2\, \bigg\langle\, \tilPsi_1\, \bigg\vert\,
       \frac{2\, \tilPhi_c\, \tilPsi_1}{h}\, \tilPhi_1\, \bigg\rangle~,
\label{eq:IP-tilPsi_1-tilPsi_2}
\end{align}
where we use the fact that $\calL_\psi$ is hermitian
and $\calL_\psi\, \tilPsi_1 = 0$.

Therefore, up to $O(\epsilon)$, the eigenvalue is given by
\begin{align}
   - \tilk_*^2
  &= \epsilon\, N/D + O(\epsilon^2)~,
\label{eq:eigenvalue_last} \\
   N
  &= 2\, \int^1_0 du~
    \left( \tilPhi_1'(u) + \tilPhi_1(0) \right)^2
  > 0~,
& & D
  := \int^1_0 du~\frac{\tilPsi_1^2(u)}{u^2}
  > 0~,
\label{eq:def-N_D}
\end{align}
and we finally obtain the superconducting coherence length as
\begin{align}
   \xi
  &= \frac{ \epsilon^{-1/2} }{\alpha(T_c)}\,
  \sqrt{ \frac{D}{\, N} }
  + O(\epsilon^{1/2})
  \propto \frac{1}{T_c}\, \left( 1 - \frac{T}{T_c} \right)^{-1/2}~.
\label{eq:coherence_length}
\end{align}
We note that since $D$ and $N$ are dimensionless quantities,
they do not depend on $T_c$ directly, but depend on $q_c$ only.
\section{diamagnetic current}
In this section, we calculate diamagnetic current
induced by a homogeneous external magnetic field
perpendicular to the surface of the superconductor.
As mentioned before, in the probe limit $e\to \infty$,
the magnetic field does not backreact to the background spacetime
(\ref{Sch-eq}).
Under the ansatz $\delta A_{y}(u, x)= b(u)\, x$
(the bulk magnetic field $F_{xy} = \p_x \delta A_{y}= b(u)$),
the equation of motion for $b(u)$ is decoupled from the other ones
for $\psi$ and $\phi$
\footnote{At the non-linear regime, the fluctuation of the gauge field
is coupled to the one of the scalar field.
This effect is important at large $x$.
As far as we are concerned with the neighborhood of the origin $x=0$,
this effect is negligible.
The non-linear effect has been considered in \cite{albashjohnson}},
and it is equivalent to Eq.(\ref{sy-eq}) for $k=0$:
\be
\label{sy'-eq}
\left( \frac{d}{du}\, h(u)\, \frac{d}{du}
- \frac{2 \tilde{\Psi}^2(u)}{u^2} \right)\, b(u) = 0~,
\ee
with the regularity boundary condition at the horizon $u=1$.

As seen in the previous section,
the equation can be solved
as a series in $\epsilon$.
Expanding $b$ as
\be
b(u) = b_{0}(u)
+ \epsilon\, b_{1}(u) + O(\epsilon^2)~,
\ee
and using Eq.(\ref{back-solution}),
we obtain equations for $b_{0}$ and $b_{1}$ as
\begin{align}
  & 0 = \odiff{}{u}\, h\, \odiff{}{u}\, b_{0}(u)~,
\label{magnetic1} \\
  & 0 = \odiff{}{u}\, h\, \odiff{}{u}\, b_{1}(u)
  - \frac{2\, \tilPsi_1^2}{u^2}\, b_{0}(u)~.
\label{magnetic2}
\end{align}
The solution of Eq.(\ref{magnetic1}) satisfying
the regularity condition is
\begin{align}
  & b_0(u) = C = (\text{const.})~.
\end{align}
So, the regularity solution of Eq.(\ref{magnetic2})
should satisfy
\begin{align}
  & \odiff{b_{1}}{u}
  = - \frac{2 C}{h(u)} \int^1_u du_0~\frac{\tilPsi_1^2(u_0)}{u_0^2}~,
\label{sI1-eq}
\end{align}
and
\begin{align}
  & b_1(u) = D - 2\, C\, \int^u_0 \frac{du_1}{h(u_1)}
    \int^1_{u_1} du_0~\frac{\tilPsi_1^2(u_0)}{u_0^2}~.
\label{eq:sol-b_1}
\end{align}
Thus, we obtain
\begin{align}
   b(u)
  &= C + \epsilon\, D
  - 2\, \epsilon\, C\, \int^u_0 \frac{du_1}{h(u_1)}
    \int^1_{u_1} du_0~\frac{\tilPsi_1^2(u_0)}{u_0^2}
  + O(\epsilon^2)
\nonumber \\
  &= B
  - 2\, \epsilon\, B\, \int^u_0 \frac{du_1}{h(u_1)}
    \int^1_{u_1} du_0~\frac{\tilPsi_1^2(u_0)}{u_0^2}
  + O(\epsilon^2)~,
\label{eq:sol-b}
\end{align}
and
\begin{align}
   \delta A_y(u, x)
  &= \delta A_y^{(0)}(x)\, \left( 1
  - 2\, \epsilon\, \int^u_0 \frac{du_1}{h(u_1)}
    \int^1_{u_1} du_0~\frac{\tilPsi_1^2(u_0)}{u_0^2} \right)
  + O(\epsilon^2)~,
\label{eq:sol-delta_A_y}
\end{align}
where we define $B := \lim_{u \to 0}\, b(u)$
and $\delta A_y^{(0)}(x) := \lim_{u \to 0}\, \delta A_y(u, x)$.

From the asymptotic behavior of $\delta A_{y}(u, x)$
near the AdS boundary,
we can read out the dual source $\delta A_y^{(0)}$ and
the thermal expectation value of the current $\Exp{J_y}$ as
\begin{align}
  & \delta A_{y}(u, x)
  = \delta A_y^{(0)}(x)
  + \frac{2\, \kappa_4^2\, e^2}{L^2}\,
  \frac{3}{4 \pi\, T}\,
    \Exp{J_y(x)}\, u + O(u^2)~.
\end{align}
From Eq.(\ref{eq:sol-delta_A_y}), we obtain
\begin{align}
   \Exp{J_y(x)}
  &= \frac{L^2}{2\, \kappa_4^2\, e^2}\,
  \frac{4 \pi\, T_c}{3}\,
    \left( - 2\, \epsilon \int^1_{0} du~\frac{\tilPsi_1^2(u)}{u^2}
    \right) \delta A_y^{(0)}(x) + O(\epsilon^2)
\nonumber \\
  &= - \frac{L^2}{2\, \kappa_4^2\, e^2}\,
  \frac{8 \pi\, T_c}{3}\,
    \left[\, L\, \Psi^{(I)}\, \right]^2\, \left(
     \frac{ \int^1_0 du~\Psi^2(u)/u^2}
          { \left[\, \Psi^{(I)}\, \right]^2 } \right)\,
  \delta A_y^{(0)}(x) + O(\epsilon^2)~,
\label{current-y}
\end{align}
where we use $L\, \Psi = \tilPsi
= \epsilon^{1/2} ( \tilPsi_1 + O(\epsilon) )$.
Because $\tilPsi_1$ (or $\Psi$ at the leading order in $\epsilon$)
is the solution of the linear equation (\ref{eq:pert-bg-EM_eq}),
we can express $\Psi(u)$ as
\begin{align}
  & \Psi(u) = \Psi^{(I)}~F(u)~,
\end{align}
where $F(u)$ is the solution of Eq.(\ref{eq:pert-bg-EM_eq})
satisfying $\lim_{u \to 0}\, F(u) =u^{\Delta_I}$ and
the regularity boundary condition at the horizon. So, 
the parenthesis in the last equation of Eq.(\ref{current-y}) depends 
on $q_c$ only, not on $\Psi^{(I)}$ and  
Eq.(\ref{current-y}) is simplified as
\begin{align}
   \Exp{J_y(x)}
  \sim - T_c\,\epsilon\, \delta A_{y}^{(0)}(x)\propto
-\,| \Exp{\calO_I} |^2\, \delta A_{y}^{(0)}(x)~.
\label{eq:current_vs_OExp}
\end{align}
Thus, the stationary current is induced only
when condensation occurs.

Interestingly, Eq.(\ref{eq:current_vs_OExp}) is very similar
to the expression expected by \GL theory.
In the theory, when the phase of the order parameter $\psi$
coupled to the U(1) gauge field $\bmA$ is constant,
the current $\bmJ$ is described by the London equation
\begin{align}
  & \bmJ = - \frac{e_*^2}{m_*}~\psi^2\, \bmA
  = - e_*\, n_s\, \bmA~,
\label{eq:London_eq}
\end{align}
where $e_*$ and $m_*$ are effective charge and mass
of the order parameter,
and $n_s$ is the superfluid number density.

While $\delta A_y^{(0)}$ in Eq.(\ref{current-y}) is an external source,
the macroscopic gauge field $\bmA$
in Eq.(\ref{eq:London_eq}) is composed of the spatial average of the
microscopic field and external field.
Since we have no dynamical photon in our holographic superconductor
model, the current does not
produce its own microscopic magnetic fields%
\footnote{We thank one of the referees for informing us
about this point}.
This means that
the external gauge field $\delta A_y^{(0)}$ is equal
to the macroscopic gauge field $\bmA$ in the AdS/CFT superconductor.
So, comparing Eq.(\ref{eq:London_eq})
with Eq.(\ref{eq:current_vs_OExp}), we obtain
the superfluid number density%
\footnote{Our superfluid number density $n_s$
is equal to the one obtained from the electric conductivity
$\Im[\sigma(\omega)]$ by Ref.~\cite{HHH2008}.}
$n_s$ which behaves as
\begin{align}
  & n_s \sim \epsilon\, T_c \sim T_c-T~,
\label{magnetic penetration depth}
\end{align}
near the critical temperature.
\section{Conclusions and Discussion}
We have investigated linear fluctuations of the scalar field solution
in the holographic model of a superconductor in Ref.\cite{HHH2008}
under the probe limit, where the fluctuations do not backreact
on the geometry.
By solving analytically the linearized equations
with only spatial momentum along one spatial coordinate
of the AdS boundary,
we find that the superconducting coherence length $\xi$
diverges at the critical temperature $T_c$
as $\xi \sim (1 - T/T_c)^{-1/2}/T_c$.
We also find diamagnetic current induced
by an external small homogeneous magnetic field.
The current is proportional to the external gauge field
and goes to zero as $T_c-T$ at the critical temperature.
These results are in agreement with the behaviors predicted
by \GL theory and Eq.~(\ref{eq:London_eq}) is the London equation
in the AdS/CFT superconductor.

If we would have dynamical photon, then
according to \GL theory,
the magnetic penetration depth $\lambda$ were related
to the superfluid density $n_s$ as
\begin{align}
  & \lambda \sim 1/\sqrt{n_s}~.
\label{eq:lambda}
\end{align}
In \GL theory, the coefficient $\kappa=\lambda/\xi$ classifies
the superconductors into two types, i.e. $\kappa<1/\sqrt{2}$
for type I superconductors and $\kappa>1/\sqrt{2}$
for type II superconductors.
Using Eq.(\ref{eq:lambda}) formally,
from Eqs.(\ref{eq:coherence_length})
and (\ref{magnetic penetration depth}) we obtain 
\be
\kappa=\frac{\lambda}{\xi}\sim T_c^{1/2}.
\ee
This may suggest that for a sufficiently small critical temperature
$T_c$, the AdS/CFT superconductor behaves as type I,
while for a sufficiently large critical temperature $T_c$,
it behaves as type II. This simple classification calls for further
investigation. \\\\
{\bf Note added:} After having submitted this article,
we learned of a work by S.A. Hartnoll, C.P. Herzog, and G.T. Horowitz,
which argued that AdS/CFT superconductor should be type II
\cite{HHH-ver2}.
We also learned of a work by G.T. Horowitz and M.M. Roberts,
which argued the dependence of the AdS/CFT
superconductor on the scalar field mass~\cite{ref:HorowitzRoberts}.
As easily seen in the derivation of $\xi$ in section III,
the mass dependence only appears
via the background scalar field solution $\tilde{\Psi}_1$
of the differential equation~(\ref{eq:bg-scalar_eq}).
Since ${\cal L}_\psi$ is still hermitian for the general mass
satisfying Breitenlohner-Freedman bound $L^2m^2 >L^2m^2_{\text{BF}}
= - 9/4$, we can extend our calculation to the general mass case.
We wish to thank one of the referees for helpful suggestions
about this extension.
\begin{acknowledgments}
We would like to thank J. Koga, M. Natsuume, and Y. Shibusa
for discussions. We are grateful to J. Goryo for comments.
\end{acknowledgments}

\end{document}